\documentclass[]{jpsj3}
\usepackage{graphicx}% Include figure files
\usepackage{amsmath,amssymb,braket,amsfonts}
\usepackage{dcolumn}% Align table columns on decimal point
\usepackage{bm}% bold math
\usepackage{color}

\title{Frustration-induced quantum spin liquid behavior in the $s=$1/2 random-bond Heisenberg antiferromagnet on the zigzag chain}

\author{
Kazuki Uematsu$^{1}$, Toshiya Hikihara$^{2}$, and Hikaru Kawamura$^{3}$\thanks{E-mail:h.kawamura.handai@gmail.com}
}

\inst{
	$^1$Department of Earth and Space Science, Graduate School of Science, Osaka University, Toyonaka, Osaka 560-0043, Japan\\
	$^2$Faculty of Science and Technology, Gunma University, Kiryu, Gunma 376-8515, Japan\\
	$^3$Molecular Photoscience Research Center, Kobe University, Kobe, Hyogo 657-8501, Japan
}

\abst{
Recent studies have revealed that the randomness-induced quantum spin liquid (QSL)-like state is stabilized in certain frustrated quantum magnets in two and three dimensions. In order to clarify the nature of this gapless QSL-like state, we investigate both zero- and finite-temperature properties of the random-bond one-dimensional (1D) $s=1/2$ Heisenberg model with the competing nearest-neighbor and next-nearest-neighbor antiferromagnetic interactions, $J_1$ and $J_2$, by means of the exact diagonalization, density-matrix renormalization-group and Hams--de Raedt methods. We find that, on increasing the frustration $J_2$, the gapless nonmagnetic state stabilized in the unfrustrated model with $J_2=0$, the {\it unfrustrated\/} random-singlet (RS) state, exhibits a phase transition into different gapless nonmagnetic state, the {\it frustrated\/} RS state. This frustrated RS state in 1D has properties quite similar to the randomness-induced QSL-like state recently identified in 2D and 3D frustrated magnets exhibiting the $T$-linear low-temperature ($T$) specific heat, while the unfrustrated RS state is more or less specific to the unfrustrated 1D system exhibiting the $\sim 1/|\log T|^3$ low-$T$ specific heat. Universal features and the robustness against perturbations of the frustrated RS state are emphasized.
}

\kword{frustration, quantum spin liquid, randomness, zigzag chain, random-singlet phase}

\begin{document}
\maketitle

\section{Introduction}
\label{sec:intro}
The quantum spin liquid (QSL), the magnetically disordered state of quantum magnets not accompanied by any spontaneous symmetry breaking down to low temperatures, has attracted much attention as a novel state of magnets since the earlier proposal of the resonating-valence bond (RVB) state by Anderson \cite{AndersonRVB}. While the proposal has long been limited to the theoretical one, many candidate materials have been reported in this century \cite{QSLreview-Balents,QSLreview-Zhou-Kanoda-Ng,KawamuraUematsu}. Many of them are {\it frustrated\/} quantum magnets such as triangular \cite{ET-Kanoda,ET-Kawamoto} and kagome \cite{Shores} magnets, which include not only the geometrically frustrated magnets, but also the other types of frustrated magnets bearing the competition between the nearest-neighbor (NN) and further-neighbor interactions \cite{Mustonen} or the one between the two-body and multi-body interactions \cite{3He3He-Ishida}, etc. Most of them are two-dimensional (2D) magnets with enhanced fluctuations, but a few three-dimensional (3D) examples \cite{Attfield} are also reported.

 The physical origin of these QSL states has hotly been debated theoretically. Though most of these approaches presume that the QSL property is an attribute of the clean system, it has recently been suggested that the randomness could often be essential in stabilizing the QSL in 2D and 3D quantum magnets \cite{Singh,Watanabe,Kawamura,Shimokawa,Uematsu,Savary,Kimchi,Uematsu2,Kimchi2,Liu-Sandvik,Wu,Uematsu3,KawamuraUematsu}. Especially, one of the present authors (H.K.) and collaborators have demonstrated by using the exact diagonalization (ED) method that the introduction of randomness to {\it frustrated} magnets universally induces the gapless QSL-like state in a variety of $s=1/2$ Heisenberg systems, including the triangular \cite{Watanabe,Shimokawa}, kagome \cite{Kawamura,Shimokawa}, $J_1$-$J_2$ square \cite{Uematsu2}, $J_1$-$J_2$ honeycomb \cite{Uematsu}, and even 3D pyrochlore \cite{Uematsu3} magnets, as long as they possess a certain amount of frustration and randomness, arguing that many of experimentally reported QSL candidates might indeed be such randomness-induced one. Hereafter, we call such randomness-induced QSL-like state in the frustrated magnets ``the frustrated random-singlet (RS) state''. Thermodynamically, the frustrated RS state in 2D and 3D frustrated magnets are gapless nonmagnetic state \cite{Watanabe,Kawamura,Uematsu,Uematsu2,Liu-Sandvik,Wu,Uematsu3} characterized by the temperature ($T$)-linear low-$T$ specific heat \cite{Watanabe,Kawamura,Uematsu,Uematsu2,Uematsu3}, the gapless susceptibility often with an intrinsic Curie-like tail \cite{Watanabe,Kawamura,Uematsu,Uematsu2,Uematsu3}, and the broad and gapless structure factor \cite{Kawamura,Shimokawa,Uematsu,Uematsu2,Wu,Uematsu3}.

 While the QSL candidates recently observed experimentally are mainly 2D with a few 3D examples, intensive theoretical studies for one-dimensional (1D) systems \cite{1DQM-review} were made in the last century concerning the effect of quenched randomness on the quantum state of the 1D quantum spin chain, typically for the {\it unfrustrated\/} 1D $s=1/2$ antiferromagnetic (AFM) Heisenberg chain \cite{DasguptaMa,MaDasguptaHu,Hirsch,Fisher}. The strong disorder renormalization group (SDRG) analysis introduced by Ma, Dasgupta and Hu \cite{MaDasguptaHu} turned out to be especially powerful. These SDRG studies revealed that the gapless nonmagnetic state consisting of spatially random covering of independent singlet-dimers was stabilized in the $s=1/2$ random-bond 1D Heisenberg spin chain with the AFM nereast-neighbor interaction. This gapless nonmagnetic state is characterized by the infinite-disorder fixed point (IDFP) where the SDRG analysis becomes asymptotically exact \cite{Fisher}. Such a state stabilized in the unfrustrated 1D chain was also called the ``random-singlet state''. In this paper, we call this state "the unfrustrated RS state", in order to distinguish it from the frustrated RS state. The SDRG analysis has revealed that the unfrustrated RS state is made only of singlet-dimers, while the distance between spins forming a singlet-dimer could be far leading to the randomness-averaged spin correlation falling off slowly with a power law. The low-temperature specific heat was predicted to behave as $\sim 1/|\log T|^3$ \cite{Hirsch}. Such SDRG analysis motivated the similar subsequent studies for other random systems \cite{SDRGreview2005-Igloi,SDRGreview2018-Igloi}, including the frustrated $J_1$-$J_2$ (zigzag) chain \cite{ladderSDRG-Igloi,zigzagSDRG-Yusuf,zigzagSDRG-Hoyos}, the ferromagnetic (FM)-AFM chain \cite{FMSDRG-WesterbergL,FMSDRG-WesterbergB,SDRGimprove-Hikihara}, and even the 2D and 3D frustrated spin systems \cite{2DSDRG-Igloi}.

 Then, the question one naturally asks might be what is the relation, {\it i.e.\/}, the similarity and the difference, between the unfrustrated and frustrated RS states. The two RS states look somewhat similar as both states are randomness-induced gapless nonmagnetic states. Nevertheless, a closer examination suggests that some apparent differences also exist between the unfrustrated and frustrated RS states. For example, the low-$T$ specific heat behaves as $\sim 1/|\log T|^3$ in the unfrustrated RS state in 1D  \cite{Hirsch,KawamuraUematsu}, while as $\sim T$ in the frustrated RS state in 2D and 3D \cite{KawamuraUematsu,Uematsu3}. Furthermore, while the unfrustrated RS state always consists of singlet ground state and hence $\left[ \langle \bm{S}_{tot}^2\rangle \right]=0$ ($\braket{\cdots}$ and $[\cdots]$ respectively represent the ground-state and the random average), the frustrated RS state contains a finite fraction of non-singlet ground states, and hence $\left[ \langle \bm{S}_{tot}^2\rangle \right]>0$.

 The miscroscopic character of the states also seems to differ somewhat, though both RS states are spin-singlet-based states. The unfrustrated RS state characterized by the IDFP consists exclusively of hierarchically organized singlet-dimers \cite{Fisher}, and the low-energy excitation is the singlet-to-triplet excitation which is more or less localized spatially. By contrast, while the frustrated RS state also contains hierarchically organized singlet-dimers, it also contains a significant amount of ``orphan spins'', unpaired spins mobile diffusively, as well as singlet-dimers clusters formed via the quantum-mechanical resonance between energetically degenerate singlet-dimers configurations \cite{KawamuraUematsu}. Most importantly, the low-energy excitations of the 2D frustrated RS state seem more `dynamical' than those in the 1D unfrustrated RS state in that the majority of low-energy excitations are the orphan-spin diffusion accompanied by the recombination of nearby singlet-dimers, and the formation and destruction of singlet-dimers clusters, which might be responsible for the different behavior of the low-$T$ specific heat of the two RS states. These observations suggest that the fixed point (FP) describing the furstrated RS state might be the {\it finite\/}-disorder FP rather than the IDFP, if one notices the fact that the IDFP means the state where each spin forms a unique singlet-dimer with a particular spin in the system. Ref.[\citen{Liu-Sandvik}] also notices for a certain 2D quantum spin model with the six-body interaction that the randomness-induced QSL state realized there is governed by the finite-disorder FP rather than the IDFP.

 Furthermore, the stability of the state seems to differ between the frustrated and unfrustrated RS states. Recent studies have revealed that the frustrated RS state is a highly universal state realized in a wide variety of quantum magnets with certain amount of frustration and randomness, independent of the lattice type, the details of interactions, and even the spatial dimensionality being two or three. Namely, the frustrated RS states in those models exhibit in common the $T$-linear specific heat originating from the singlet-dimers recombination \cite{Watanabe,Kawamura,Uematsu,Uematsu2,Uematsu3}, the Curie-like gapless susceptibility arising from orphan spins \cite{Watanabe,Kawamura,Uematsu,Uematsu2,Uematsu3}, and the broad structure factor reflecting the absence of characteristic energy scales \cite{Kawamura,Shimokawa,Uematsu,Uematsu2,Wu,Uematsu3}.

 Such robustness of the frustrated RS state is in apparent contrast to that of the unfrustrated RS state, which is rather fragile against weak perturbations such as the introduction of frustration \cite{zigzagSDRG-Hoyos} and the 2D (3D) coupling \cite{Laflorencie}. Recall that the unfrustrated RS state stabilized in the 1D unfrustrated AFM Heisenberg chain is destroyed by the introduction of an infinitesimal amount of frustration ($J_2/J_1$)\cite{zigzagSDRG-Hoyos}, and the unfrustrated random antiferromagnets in $d\geq2$ dimensions generically exhibit the standard AFM order even with the extremely strong randomness \cite{Laflorencie}. Such a contrast in the stability of the frustrated and unfrustrated RS states might also suggest some fundamental difference between the two RS states.

 In view of such a situation, it would be highly desirable to further clarify the relation between the frustrated RS state recently identified in frustrated 2D and 3D random systems and the unfrustrated RS state identified earlier in unfrustrated 1D random system. For this purpose, we study in the present paper the {\it frustrated\/} 1D random system, {\it i.e.\/}, the random $s=1/2$ Heisenberg chain with the competing AFM NN and next-nearest-neighbor (NNN) interactions, $J_1$ and $J_2$. For $J_2=0$, the model reduces to the well-studied unfrustrated Heisenberg chain, for which the existence of the unfrustrated RS state described by the IDFP has been established.

 When applied to the {\it frustrated} random $J_1$-$J_2$ chain \cite{ladderSDRG-Igloi,zigzagSDRG-Yusuf,zigzagSDRG-Hoyos}, it was shown that the SDRG method lead to the large-spin ground state characterized by the large-spin FP, the FP which is different from the unfrstrated RS FP, even for an infinitesimally small $J_2$, {\it i.e.\/}, $J_2/J_1\gtrsim10^{-6}$ \cite{zigzagSDRG-Hoyos}. The large-spin FP is the 1D counterpart of the spin-glass (SG) FP where the ground-state total spin $S_{tot}$ behaves as $S_{tot} \sim L^{1/2}$ ($L$ the total number of spins). The same relation holds also in the 1D FM-AFM model, a typical SG-like model with the random mixture of the FM and AFM NN interactions \cite{FMSDRG-WesterbergL,FMSDRG-WesterbergB,SDRGimprove-Hikihara}, which is believed to be governed by the large-spin FP. The SDRG method predicted that the low-$T$ specific heat of the large-spin state behaved as $C\propto T^{1/z}\log T$, where the dynamical exponent $z$ was estimated to be $1/z\lesssim0.15$ in the $J_1$-$J_2$ chain \cite{zigzagSDRG-Hoyos}, and $1/z\sim0.44$ in the FM-AFM chain \cite{FMSDRG-WesterbergL,FMSDRG-WesterbergB}. Nevertheless, the validity of the SDRG method applied to the frustrated $J_2>0$ random chain is not clear in contrast to the case of the unfrustrated random chain of $J_2=0$. Furthermore, there are rather few direct numerical calculations like the ED or density-matrix renormalization-group (DMRG) methods performed for the random $J_1$-$J_2$ chain \cite{Lavarelo}. Hence, the issue of the ground state of the random $J_1$-$J_2$ chain with non-negligible frustration still seems to deserve careful numerical examination.

 In our present study of the random $J_1$-$J_2$ chain, the ED and DMRG methods are applied to map out the phase diagram of the model, with particular attention to the question of whether the unfrustrated RS state is really destabilized by the introduction of frustration $J_2$, and if so, whether the frustration-induced state is really the large-spin state as predicted by the SDRG analysis or some other state such as the frustrated RS state. The appearance of the frustrated RS state might not be so unusual, if one recalls the insensitivity of the frustrated RS state to the spatial dimensionality $d$ \cite{Uematsu3}, at least for the cases of $d=2$ and 3. If such an insensitivity of the frustrated RS state to the dimensionality is to be extended to $d=1$, both the unfrustrated RS state and the frustrated RS state might be stabilized in the $J_1$-$J_2$ random chain with varying the frustration strength $J_2$, {\it i.e.\/}, the unfrustrated RS state for $J_2=0$ (or $J_2<J_c$) and the frustrated RS state for $J_2>0$ (or $J_2>J_c$), which are separated via a phase transition. Such an observation would unambiguously indicate that the frustrated RS state and the unfrustrated RS state are distinct quantum states.

 Indeed, we find that the frustrated RS state, instead of the large-spin state, is stabilized in the $J_2>J_c$ region. In this way, we expect that the better understanding of the ground state of the frustrated 1D $J_1$-$J_2$ model could provide some useful information even on the nature of the RS state in general, identified for the frustrated 2D and 3D systems.

 The rest of the paper is organized as follows. In Sect. \ref{sec:model}, we introduce the models analyzed in the present paper, {\it i.e.\/}, the $s=1/2$ random-bond isotropic Heisenberg model with the competing NN and NNN AFM interactions $J_1$ and $J_2$, and the $s=1/2$ random isotropic Heisenberg model with the FM and AFM NN interactions. While our main focus is on the former model, the latter model is also studied to better understand the nature of the ground state of the former model. The details of our numerical computations, {\it i.e.\/}, the ED and DMRG calculations, are explained. The results of the ED calculations are presented in Sect. \ref{sec:GSzigzag}. In Sect. \ref{sec:GSphase}, we present the ground-state phase diagram of the $J_1$-$J_2$ model in the frustration ($J_2$) versus the randomness ($\Delta$ to be defined below) plane. In Sect. \ref{sec:uRS-fRS}, we focus on the relation between the unfrustrated RS state stabilized in the weaker $J_2$ region and the frustrated RS state stabilized in the stronger $J_2$ region. In Sect. \ref{sec:fRS-LS}, we further clarify the nature of the frustrated RS state by comparing its thermodynamic properties with those of the random FM-AFM chain expected to be in the large-spin state. The results of the DMRG calculations on the spin-spin correlation function of the $J_1$-$J_2$ model and the random FM-AFM model are presented in Sect. \ref{sec:correlation}. Particular attention is paid to the spatial-decay exponent of the spin-spin correlation function. In Sect. \ref{sec:excitation}, microscopic character of the ground state and the low-lying excitations of the 1D frustrated RS state is studied via the singlet-dimer configurations. Finally, Sect. \ref{sec:summary} is devoted to summary of the results.

\section{The model and the method}
\label{sec:model}
 The model we study is mainly the random-bond $s=1/2$ Heisenberg spin chain with frustrating NN and NNN AFM interactions, $J_1$ and $J_2$. The Hamiltonian is given by
\begin{align}
\mathcal{H}=J_1\sum_{\Braket{i,j}_1}\tilde{j}_{ij}{\bm S}_i\cdot{\bm S}_j +
J_2\sum_{\Braket{i,j}_2} \tilde{j}_{ij}{\bm S}_i\cdot{\bm S}_j,
\label{eq:hamiltonianJ1J2chain}
\end{align}
where $\bm{S}_i=(S_i^x, S_i^y, S_i^z)$ is the $s=1/2$ spin operator at the $i$-th site of the 1D chain, the sums $\Braket{i,j}_1$ and $\Braket{i,j}_2$ are taken over all NN and NNN pairs, and $\tilde{j}_{ij}$ is the independent random variable obeying the same uniform distribution between $[1-\Delta,1+\Delta]$ with $0\leq\Delta\leq1$. We put $J_1=1$ and $J_2/J_1=J_2>0$. The parameter $J_2$ then represents the degree of frustration borne by the competition of $J_1$ and $J_2$. The parameter $\Delta$ represents the extent of the randomness. For simplicity, we take the extent of the randomness $\Delta$ to be common between $J_1$ and $J_2$. Note that, by tuning the parameters $\Delta$ and $J_2$, we can control the degrees of both the randomness and the frustration independently.

 Furthermore, in order to better understand the data of the $J_1$-$J_2$ chain, we also study the random FM-AFM Heisenberg chain whose Hamiltonian is given by
\begin{align}
\mathcal{H}=J_1\sum_{\Braket{i,j}_1}\tilde{j}^\prime_{ij}{\bm S}_i\cdot{\bm S}_j,
\label{eq:hamiltonianFMAFMchain}
\end{align}
where $\tilde{j}^\prime_{ij}$ is the independent random variable obeying the uniform distribution between $[-1,1]$, similarly to the typical SG model.

 The ground-state properties of the $J_1$-$J_2$ chain and the FM-AFM chain are studied both by the ED Lanczos method and the DMRG method. We treat finite-size clusters with the total number of spins $L$ up to $L\leq 32$ with periodic boundary conditions (BC) in the ED calculation ($L$ is taken to be a multiple of 4 with $8\leq L\leq 32$), and $L=24,32,48$, and 64 with open BC in the DMRG calculation. The numbers of independent bond realizations $N$ used in the sample averaging are $N=999$ for $L=8$--$28$ and 120 for $L=32$ for the order parameter and the spin gap in the ED calculation, and $N=100$ for all $L$ in the DMRG calculation. In the DMRG calculation, detailed checks of the data convergence and the data consistency are performed to avoid problems caused by the massive degeneracy near  the ground state (see SIV of supplemental materials for more details \cite{suppl}). Error bars are estimated from sample-to-sample fluctuations. %The computed ground-state properties are mutually consistent between the ED and DMRG calculations up to the maximum size $N=32$ where both data are available.

 The finite-temperature properties of the $J_1$-$J_2$ chain and the FM-AFM chain are computed by the Hams--de Raedt method \cite{HamsRaedt,cTPQ}. The computation is performed with the averaging made over $N_v$ initial vectors and $N$ independent bond realizations. In the case of the $J_1$-$J_2$ chain, the employed $(N_v,N)$ values are $(70,100)$ for $L=20$, $(30,100)$ for $L=24$, and $(20,25)$ for $L=28$ for the frustrated case of $J_2>0$, while $(100, 100)$ for $L=20$ and $24$, $(45, 100)$ for $L=28$, and $(10, 25)$ for $L=32$ for the unfrustrated case of $J_2=0$, respectively. In the case of the FM-AFM chain, they are $(70,100)$ for $L=20$, $(30,100)$ for $L=24$, and $(20,100)$ for $L=28$. Error bars of physical quantities are estimated from the scattering over both samples and initial states by using the bootstrap method.

\section{The exact-diagonalization study}
\label{sec:GSzigzag}
 In this section, we present the results of our ED calculations on the random $J_1$-$J_2$ chain both at $T=0$ and at $T>0$, including the unfrustrated case of $J_2=0$. We also study the properties of the random FM-AFM chain to clarify the properties of the $J_1$-$J_2$ chain.

%\subsection{IIIA. The phase diagram}
\subsection{The phase diagram}
\label{sec:GSphase}
\begin{figure*}
 \raisebox{1.5cm}{\includegraphics[width=12cm]{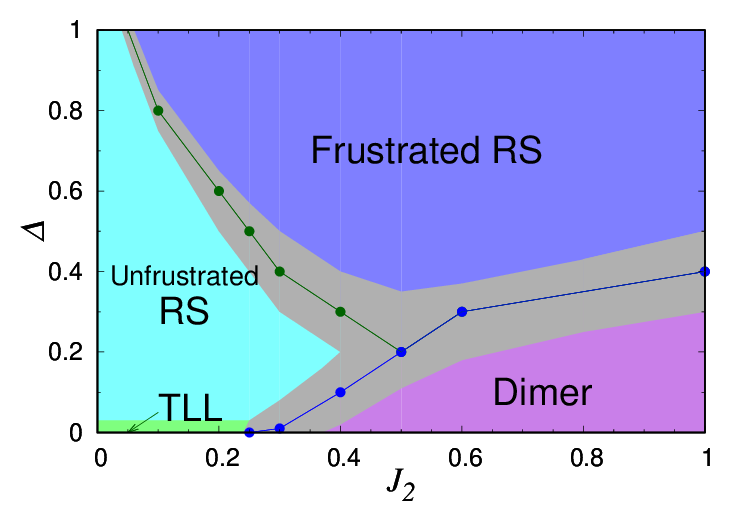}}
	\caption{
(Color online) Ground-state phase diagram of the $s=1/2$ random-bond $J_1$-$J_2$ Heisenberg spin chain in the frustration ($J_2$) versus the randomness  ($\Delta$) plane. ``TLL'' and ``Dimer'' represent the gapless Tomonaga-Luttinger liquid and the dimerized state with a finite spin gap, respectively. The blue points denote the transition points estimated from the spin gap, while the green points denote those estimated from the ratio of the samples with non-singlet ground states of $\braket{\bm{S}_{tot}^2}>0$, $R$. Grey regions represent the uncertainty of the phase boundary.
	} 
	\label{fig:phase-zigzag}
\end{figure*}

Before presenting the detailed data, we first show the ground-state phase diagram of the $J_1$-$J_2$ chain in Fig. \ref{fig:phase-zigzag} in the frustration ($J_2$) versus the randomness ($\Delta$) plane. The $\Delta=0$ line corresponds to the regular model studied in previous works \cite{zigzag-Tonegawa,zigzag-White,zigzag-Chen}.
%, where the dimerization was observed for $J_2\gtrsim0.24$.

In the phase diagram shown in Fig. \ref{fig:phase-zigzag}, four distinct phases are identified. Two of them, {\it i.e.}, the Tomonaga-Luttinger liquid (TLL) and the gapped dimerized phase have already been identified in the regular model. Though the dimerized phase has a finite width along the $\Delta$ axis in the phase diagram of Fig. \ref{fig:phase-zigzag}, this might be the finite-size effect since the dimerized phase accompanied by spontaneous symmetry breaking was reported to be unstable against the introduction of randomness \cite{dimer-weakrandom}. The unfrustrated RS phase stabilized in the weaker-$J_2$ region is a well-known phase extensively investigated in the SDRG study \cite{Fisher}, which is characterized by the singlet ground state $\braket{{\bm S}_{tot}^2}=0$ and the low-temperature specific heat $C\propto 1/|\log T|^3$ \cite{Fisher,Hirsch}. The SDRG analysis indicated that the unfrustrated RS state was unstable against the introduction of very weak $J_2\lesssim10^{-6}$ \cite{zigzagSDRG-Hoyos}. 
%where the unfrustrated RS state is unstable to the introduction of very weak $J_2\lesssim10^{-6}$ \cite{zigzagSDRG-Hoyos}, 
In our computation, the unfrustrated RS phase seems to spread over a wider finite-$J_2$ region than that predicted by the SDRG analysis, but this would probably be a finite-size effect due to the small sizes of our ED calculation.

 According to our present analysis, the upper-right phase turns out to be the {\it frustrated} RS phase as will be later detailed, whose properties are quite similar to the one stabilized in the $d\geq 2$ frustrated spin systems. The SDRG studies reported that the large-spin FP dominated the relevant part of the phase diagram \cite{zigzagSDRG-Hoyos}. We observe, however, that some of the properties of the random $J_1$-$J_2$ chain, {\it e.g.\/}, the spin freezing parameter and the low-temperature specific heat, are qualitatively different from those of the random FM-AFM spin chain which is expected to be governed by the large-spin FP: See Sect. \ref{sec:fRS-LS} below for more details.

Now, we describe how we draw the phase diagram shown in Fig. \ref{fig:phase-zigzag}. The phase boundary between the unfrustrated RS phase and the frustrated RS phase, represented by green points in Fig. \ref{fig:phase-zigzag}, is determined from the ratio of the non-singlet ground-state samples with $\braket{{\bm S}_{tot}^2}>0$, $R$, by using the property that $R=0$ in the unfrustrated RS state while $R>0$ in the frustrated RS state. The details will be given in Sect. \ref{sec:uRS-fRS} below, with some additional information also given in SII of supplemental materials \cite{suppl}.

The phase boundary between the dimer phase and the two RS phases, represented by blue points in Fig. \ref{fig:phase-zigzag}, are determined by the spin gap $\Delta E$ by using the properties that $\Delta E>0$ in the dimer phase while $\Delta E=0$ in the two RS phases. Details are given in SII of supplemental materials \cite{suppl}.

As mentioned, we conclude that the gapless nonmagnetic phase in the upper-right part of the phase diagram is the frustrated RS phase, rather than the large-spin phase as suggested by the SDRG analysis \cite{zigzagSDRG-Hoyos}. Since both the frustrated RS phase and the large-spin phase are gapless nonmagnetic states with $R>0$, we finally need to examine the finite-temperature properties, the low-temperature specific heat in particular, to discriminate the two phases. The details of the distinction between the frustrated RS phase and the large-spin phase will be given in Sect. \ref{sec:fRS-LS} below. In our identification, we shall make a comparative study of the $J_1$-$J_2$ model in the relevant parameter range, together with the random FM-AFM chain which is believed to exhibit the large-spin FP behavior, by highlighting the difference between the two models. Additional information about the finite-temperature properties of these models is given in SIII of supplemental materials \cite{suppl}.

%\subsection{IIIB. The unfrustrated random-singlet phase v.s. the frustrated random-singlet phase}
\subsection{The unfrustrated random-singlet phase v.s. the frustrated random-singlet phase}
\label{sec:uRS-fRS}

In this subsection, we focus on the unfrustrated RS phase and the frustrated RS phase, their relation and distinction in particular, and show how we determine the associated phase boundary represented by green points in the phase diagram of Fig. \ref{fig:phase-zigzag}. As mentioned in the previous section, the unfrustrated RS state has been established in the presence of randomness $\Delta>0$ at $J_2=0$. The state is nonmagnetic and gapless, similarly to the frustrated RS state identified in 2D and 3D, and to the SG-like large-spin phase. Meanwhile, in contrast to the frustrated RS phase and the large-spin phase where the rate of the $\braket{{\bm S}_{tot}^2}>0$ samples, $R$, is nonzero, it has been shown on the basis of the Marshall-Lieb-Mattis theorem that $R=0$ in the unfrustrated RS state of $J_2=0$ \cite{MLMtheorem}. Hence, the ratio $R$ might be used to detect the phase transition from the unfrustrated RS phase to another phase.

\begin{figure*}
	\includegraphics[width=8cm]{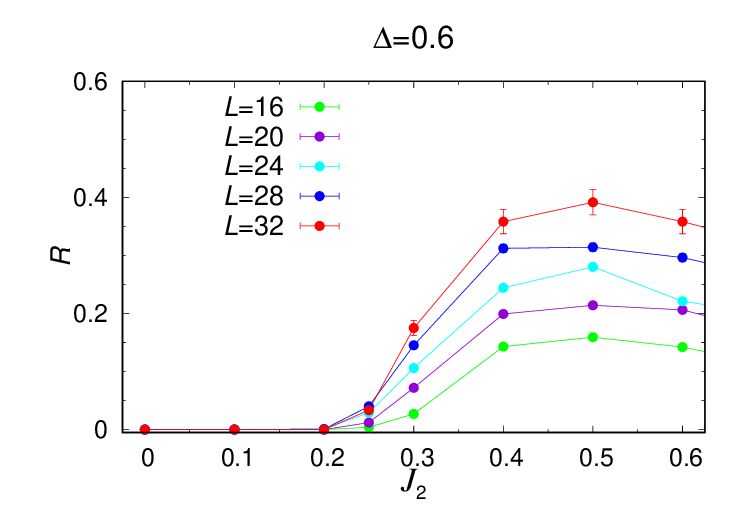}	
\vskip 0.5cm
	\caption{
		(Color online) The ratio of the samples with non-singlet ground states of $\braket{{\bm S}_{tot}^2}>0$, $R$, plotted versus $J_2$ for $\Delta=0.6$. 
	}
	\label{fig:orderchain3}
\end{figure*}

In Fig. \ref{fig:orderchain3}, we plot the ratio $R$ v.s. the frustration strength $J_2$ in the case of randomness $\Delta=0.6$ for the random $J_1$-$J_2$ chain. As can be seen from the figure, $R$ takes a nonzero value when $J_2$ exceeds $\sim0.2$, apparently exhibiting a maximum with increasing $J_2$ at around $J_2\simeq 0.5$. As mentioned, the SDRG suggests that even an infinitesimally small $J_2\sim10^{-6}$ might destabilize the unfrustrated RS phase, whereas our estimate of the critical value $J_{2c}\sim0.2$ is considerably larger. Since $R$ tends to get larger with increasing the system size $L$, and a vanishing $R$ might eventually change into a small nonzero value on increasing the number of samples $N$, finite-size and finite-sampling effects tend to overestimate the $J_{2c}$-value somewhat. Yet, although precisely locating the $J_{2c}$-value is a numerically difficult task, our present data suggest that $J_{2c}$ might significantly be greater than $\sim10^{-6}$, the value suggested from the SDRG analysis \cite{zigzagSDRG-Hoyos}. Anyway, the behavior of $R$ indicates that, on increasing the frustration $J_2$, the unfrustrated RS phase of $R=0$ exhibits a phase transition into another phase of $R>0$.

\begin{figure*}
		\includegraphics[clip,width=8cm]{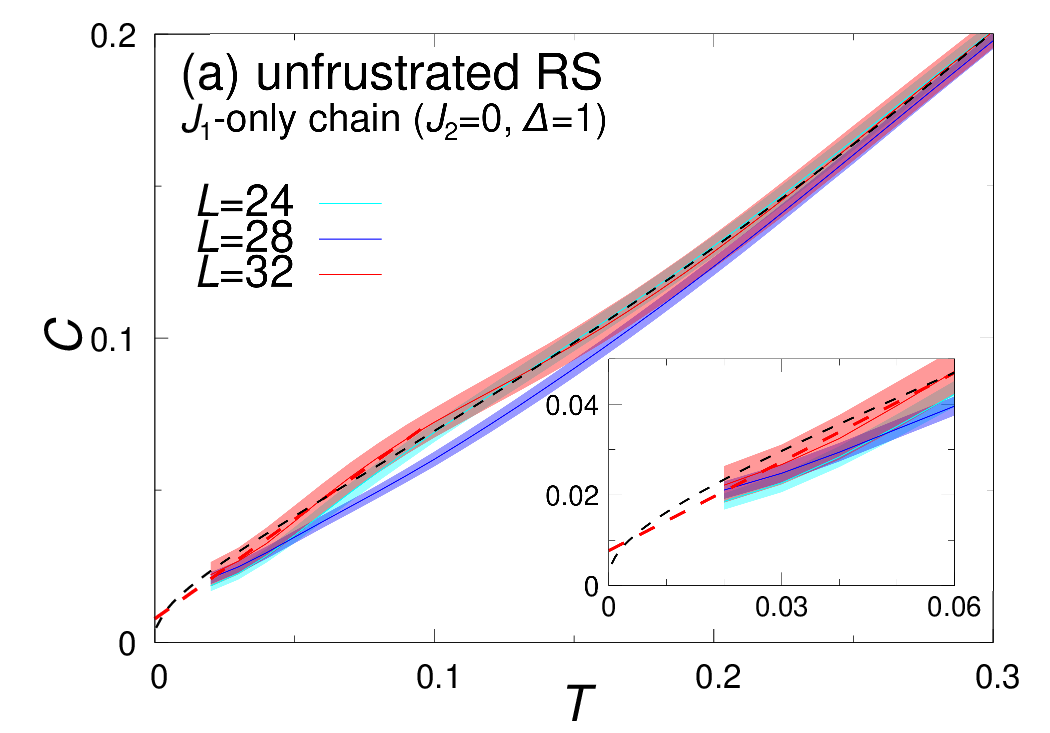}
		\includegraphics[clip,width=8cm]{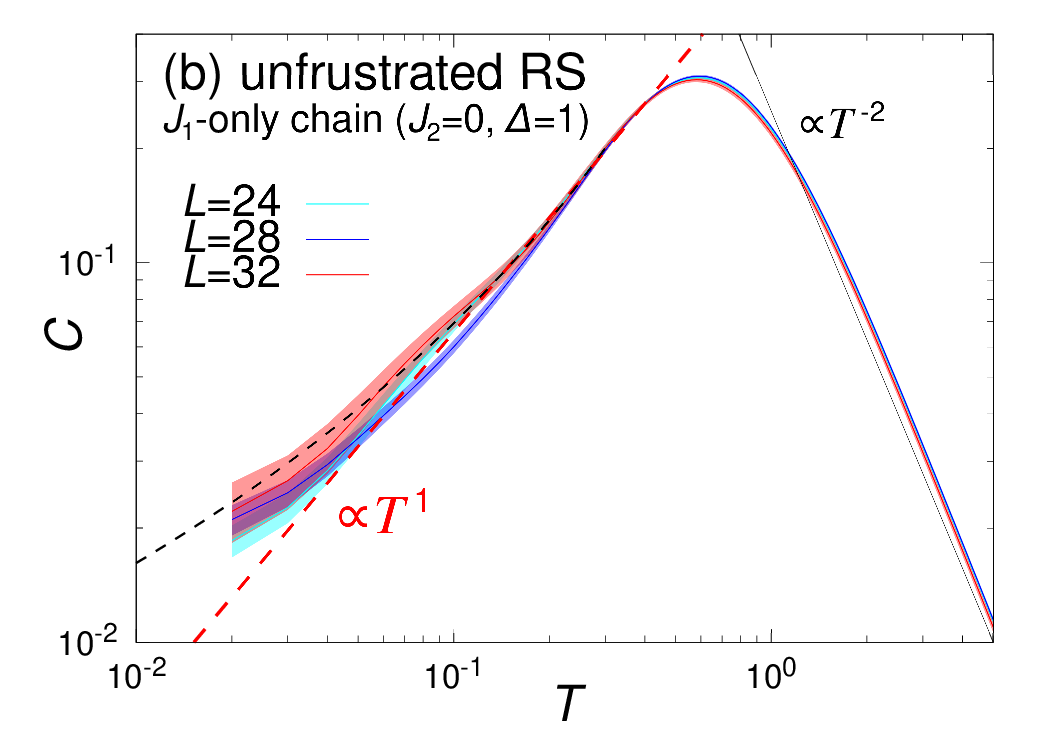}
	\caption{
		(Color online) The temperature and size dependence of the specific heat per spin $C$ of the random $J_1$-$J_2$ chain of $\Delta=1$ in the unfrustrated ($J_2=0$) case. Figure (a) shows a linear plot, together with a magnified view of the low-$T$ region in the inset, while Fig. (b) shows a log-log plot. The dashed line in Fig. (a) is a linear fit of the data at $T\leq0.1$, while the dashed line in Fig. (b) with the slope unity is the guide to the eye. The dashed black lines are the fits based on the SDRG from $\sim 1/|\log (T/T_0)|^3$ with $T_0=3.98$. The solid black line in Fig. (b) is the high-$T$ expansion result.
	}
	\label{fig:chainheat}
\end{figure*}
\begin{figure*}
		\includegraphics[clip,width=8cm]{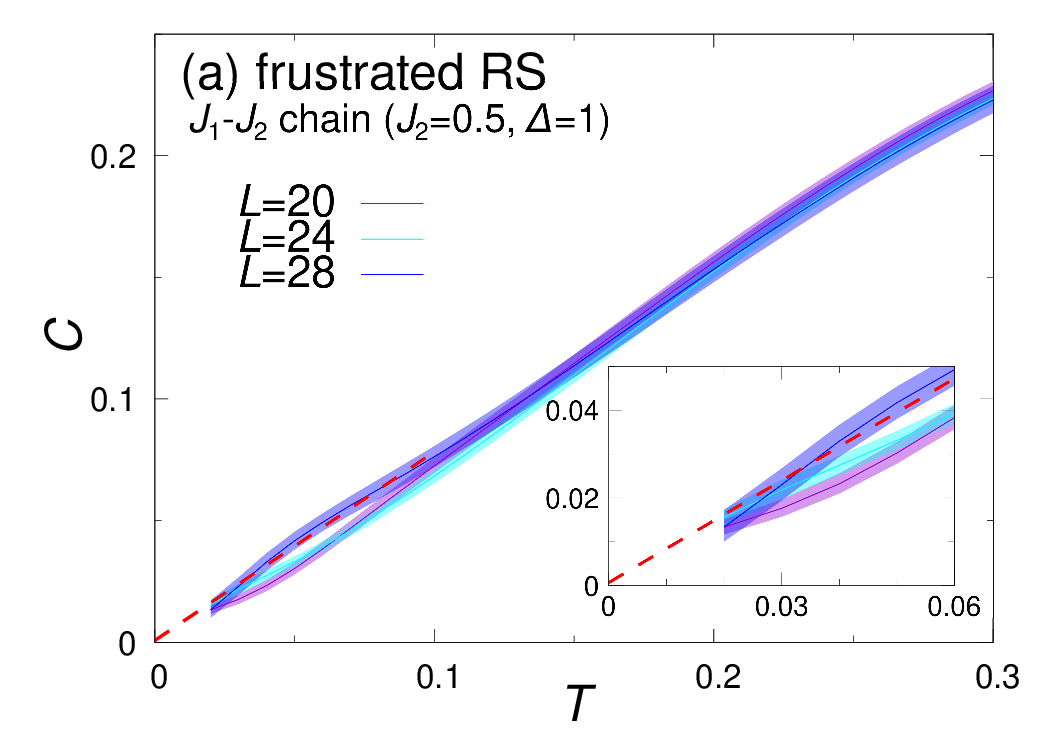}
		\includegraphics[clip,width=8cm]{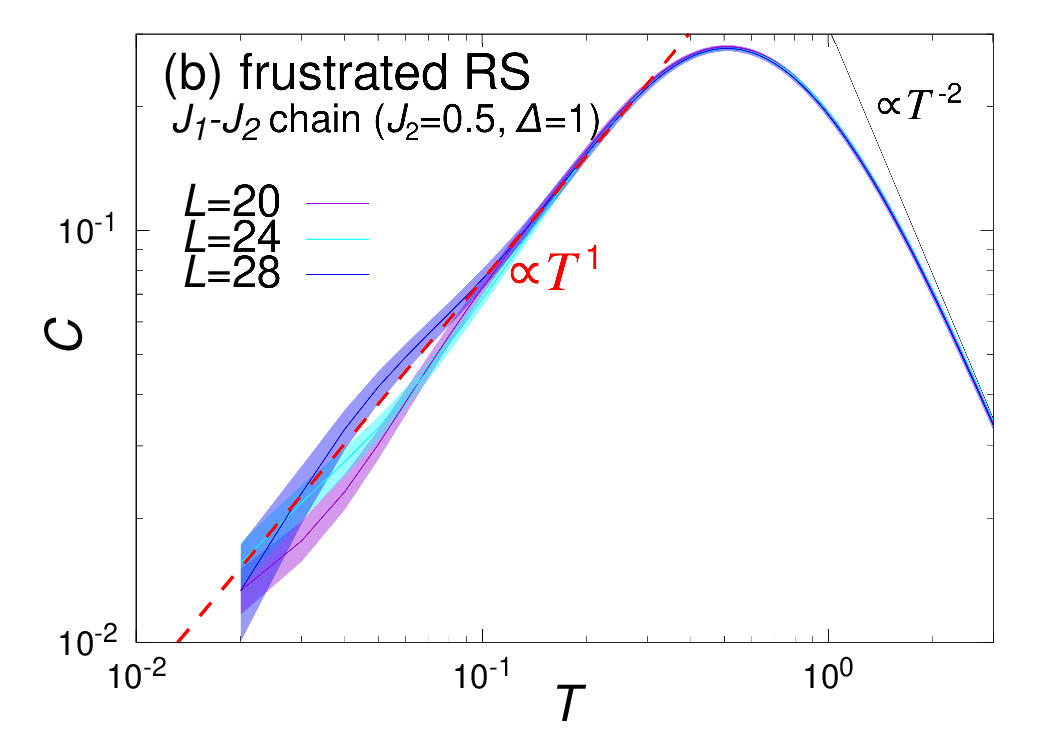}
	\caption{
		(Color online) The temperature and size dependence of the specific heat per spin $C$ of the random $J_1$-$J_2$ chain of $\Delta=1$ in the frustrated ($J_2=0.5$) case. Figure (a) shows a linear plot, together with a magnified view of the low-$T$ region in the inset, while Fig. (b) shows a log-log plot. The dashed red line in Fig. (a) is a linear fit of the data at $T\leq0.1$, while the dashed red line in Fig. (b) with the slope unity is the guide to the eye. The solid black line in Fig. (b) is the high-$T$ expansion result.
	}
	\label{fig:chainheat1}
\end{figure*}

 Further evidence of the $J_2$-induced phase transition can also be obtained from the behavior of the low-temperature specific heat $C$. In Fig. \ref{fig:chainheat}, we show the temperature dependence of the specific heat of the random $J_1$-$J_2$ chain of $\Delta=1$ for the unfrustrated case of $J_2=0$ lying in the unfrustrated RS state. In the unfrustrated RS state, it has been established that the low-temperature specific heat behaves as $C\sim 1/|\log T|^3$ \cite{Hirsch}. The computed specific heat exhibits the behavior consistent with this expectation. Namely, the linear extrapolation of the data in Fig. \ref{fig:chainheat} (a) yields a nonzero $C(T=0)>0$ value apparently violating the third law of thermodynamics, suggesting the asymptotic behavior falling abruptly toward $T=0$, which is also supported by the log-log plot of $C$ shown in Fig. \ref{fig:chainheat} (b) where the tendency of the upward bending is discernible as $T$ is lowered toward $T=0$. Indeed, the fit to the SDRG form $\sim 1/|\log(T/T_0)|^3$ yields a reasonable fit with $T_0\simeq 4.0$ as shown in Fig. \ref{fig:chainheat}. Finite-size effects turn out to be relatively minor, the data of different $L$ almost overlapping within the error bars. Especially, the features mentioned above are robust against the sizes.

 Such a behavior is in contrast to that of the frustrated $J_1$-$J_2$ model of $J_2=0.5$ and $\Delta=1$ shown in Fig. \ref{fig:chainheat1}. In fact, as can be seen from Figs. \ref{fig:chainheat1} (a) and (b), the $T$-linear behavior is observed in the frustrated case of $J_2=0.5$, quite similar to the ones observed in the frustrated RS phase in 2D and 3D \cite{KawamuraUematsu,Watanabe,Kawamura,Uematsu,Uematsu2,Uematsu3}. Although some amount of finite-size effects are appreciable, the near $T$-linear behavior with a vanishing tangent is observed rather robustly in the low-$T$ range, strongly suggesting that this feature persists in the thermodynamic limit. Such an observation indicates that the $J_2$-induced onmagnetic gapless state is the different phase from the unfrustrated RS phase (finally to be identified as the frustrated RS phase). Note that the obtained $T$-linear behavior in Fig. \ref{fig:chainheat1} is in sharp contrast also to that suggested from the SDRG analysis, $C\sim T^{1/z}\log T$ with $1/z\lesssim 0.44$ \cite{FMSDRG-WesterbergL,FMSDRG-WesterbergB,zigzagSDRG-Hoyos}. We shall discuss the point in more detail in the next subsection, also in reference to the large-spin phase.

 We also compute the entropy for the cases of $J_2=0$, 0.2 and 0.5.  For all cases studied, the entropy per spin exhibits relatively minor size dependence, taking values around $\sim 0.04$ at the lowest temperature studied $T\sim 0.02$. In the $J_2>0$ case, since our $R$ data indicate that a significant part of the samples possess non-singlet ground states,  a nonzero entropy of this order seems reasonable for finite-size systems. Meanwhile, due to the limitation of the available system size and the precise information about the asymptotic size dependence, and also due to the lack of quantitative information of the low-lying excitations contributing to the $T\rightarrow 0$ entropy, the reliable extrapolation to the $L\rightarrow \infty$ and $T\rightarrow 0$ limit is hard so that we cannot make a definitive statement about the existence or nonexistence of the residual entropy. 

 The uniform susceptibility is also computed for the cases of $J_2=0,\ 0.2$ and $0.5$, and the results are given in Fig.S6 of supplemental materials,\cite{suppl} together with the one for the random FM-AFM chain. The computed susceptibilities exhibit in common the Curie-like divergent behavior in the $T\rightarrow 0$ limit, similar to the ones observed in the frustrated RS phase in higher-dimensional systems \cite{Watanabe,Kawamura,KawamuraUematsu}. In addition, as can be seen from the insets of Figs. S6(a) and S6(b), the susceptibility exhibits a changeover from the high-$T$ Curie-like behavior to the low-$T$ Curie-like behavior, accompanied by a flattening behavior in the crossover temperature range of $0.3\lesssim T\lesssim 1$. In the frustrated RS phase whose low-$T$ specific heat exhibits the $T$-linear behavior $\sim \gamma T$, such a flattening behavior of the susceptibility enables one to estimate the so-called Wilson ratio $R_W$, yielding $R_W\sim 2.5$. The obtained value of $R_W$ is a bit larger, but rather close to the corresponding values of the frustrated RS state in higher dimensions, $R_W\sim 2$ \cite{D-thesis}. Further details are given in SIII of supplemental materials.\cite{suppl}

%\subsection{IIIC. The frustrated random-singlet phase v.s. the large-spin phase}
\subsection{The frustrated random-singlet phase v.s. the large-spin phase}
\label{sec:fRS-LS}

\begin{figure*}
		\includegraphics[width=8cm]{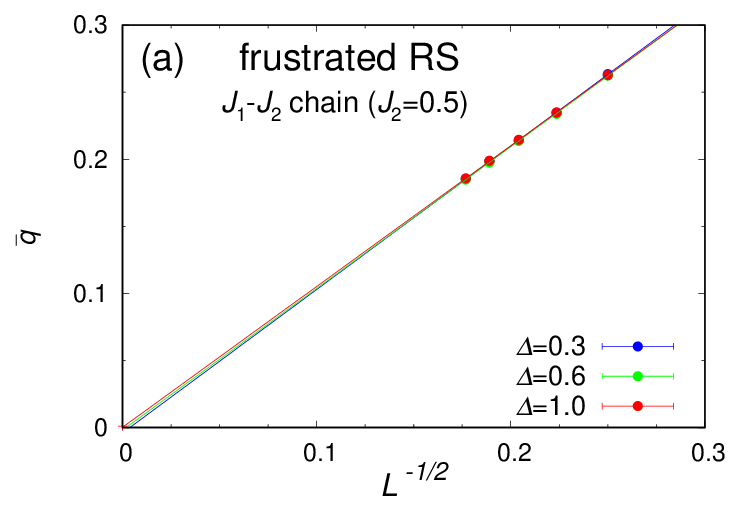}
		\includegraphics[width=8cm]{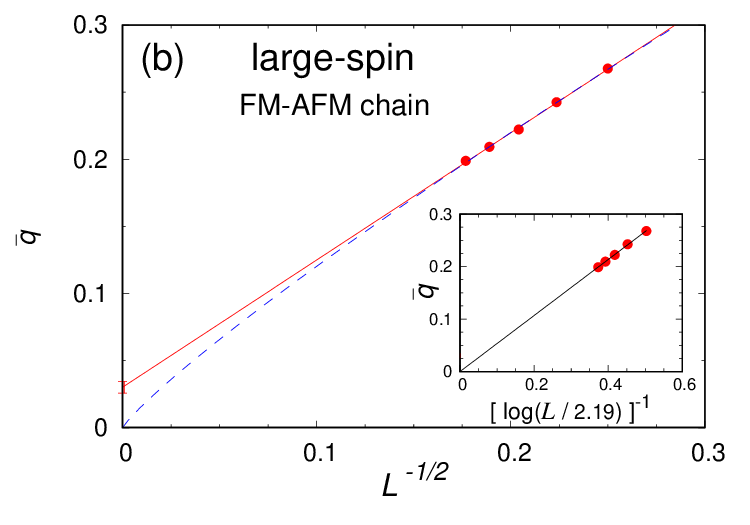}
\vskip 0.5cm
	\caption{
 (Color online) The spin freezing parameter $\bar{q}$ plotted versus $1/\sqrt{L}$. Figure (a) represents $\bar{q}$ of the random $J_1$-$J_2$ chain of $J_2=0.5$ for various values of $\Delta$, while (b) represents that of the random FM-AFM model. The solid lines are linear fits to all the data points of $L$, while the dashed blue line in (b) is the power-law fit $\sim L^{-0.44}$. In the inset of Fig. (b), the same data are plotted versus $1/\log(L/2.19)$.
	}
	\label{fig:orderchain1}
\end{figure*}

In this subsection, we identify the nature of the phase stabilized in the upper-right region in the phase diagram of Fig. \ref{fig:phase-zigzag}. In addition to the random $J_1$-$J_2$ chain, we also consider in this subsection the random FM-AFM chain as a typical 1D model realizing the large-spin (SG-like) phase, to examine whether the randomness-induced phase stabilized in the frustrated $J_1$-$J_2$ ($J_2>0$) model is the large-spin phase or not.

Figure \ref{fig:orderchain1} shows the spin freezing parameter $\bar{q}$ of (a) the random $J_1$-$J_2$ chain of $J_2=0.5$ and (b) the random FM-AFM chain. The spin-freezing parameter $\bar{q}$ is defined by
\begin{align}
&\bar{q}=\sqrt{q^{(2)}}, 
&q^{(2)}= \frac{1}{L^2}
\sum_{i,j}\left[\Braket{\bm{S}_i\cdot\bm{S}_j}^2\right].
\label{eq:qbar}
\end{align}

As will be shown in Sect. IV below, our DMRG calculation indicates that the spin correlation function $[|\braket{\bm{S}_i\cdot\bm{S}_{i+r}}|]$ of the frustrated RS state falls with a power-law $\sim r^{-\rho}$ with the exponent $1.3\lesssim \rho \lesssim 1.6$. Furthermore, the analysis in SI of supplemental materials \cite{suppl} indicates that, if the inequality $\rho>\frac{1}{2}$ holds as indicated by our DMRG analysis, the size dependence of the spin freezing parameter $\bar{q}$ should be given by $\bar{q}=\bar{q}_{\infty}+c_1'/\sqrt{L}$, instead of the spin-wave form $\bar{q}=\bar{q}_{\infty}+c_1/L$.

 Thus, in Fig. \ref{fig:orderchain1} (a) we plot $\bar{q}$ for various $L$ versus $1/\sqrt{L}$. The data lie on a straight line as expected, and are extrapolated to zero for all $\Delta$ shown here, indicating the absence of any kind of magnetic long-range order including the SG order.

 In Fig. \ref{fig:orderchain1} (b), on the other hand, $\bar{q}$ of the random FM-AFM chain plotted versus $1/\sqrt{L}$ exhibits somewhat different behavior, apparently extrapolated to a positive value suggesting a difference from the ground state of the frustrated $J_1$-$J_2$ chain. Of course, the positive $\bar{q}$-value suggested from Fig. \ref{fig:orderchain1} (b) is a spurious one originating from the slower decay of the spin correlation function \cite{SDRGimprove-Hikihara}. Indeed, the analysis of our DMRG calculation in Sect.IV below suggests the slower decay with the decay exponent $\rho$ close to, perhaps slightly smaller than 0.5. As shown in the analysis of supplemental materials \cite{suppl}, if $\rho$ is smaller than $d/2=0.5$, $\bar{q}$ should decay as $\sim L^{-\rho}$, not as $L^{-1/2}$. Then, we perform the fit based on the power-law form $\bar{q}=AL^{-\rho}$ with $\rho$ and $A$ the fitting parameters. The best fit is obtained for $\rho=0.44$ as shown in Fig. 5 (b), consistently with the DMRG results in Sect. IV below. We note that our present data of the random FM-AFM chain are also consistent with the extremely slow logarithmic decay \cite{SDRGimprove-Hikihara}, as can be seen from the inset of Fig. \ref{fig:orderchain1} (b) where we plot $\bar{q}$ versus $1/\log(L/2.19)$. Anyway, the different behaviors of $\bar{q}$ between the random $J_1$-$J_2$ chain and the random FM-AFM chain provide a supporting evidence that the phase of the random $J_1$-$J_2$ chain with non-negligible $J_2$ is not characterized by the large-spin FP.

\begin{figure*}
		\includegraphics[width=8cm]{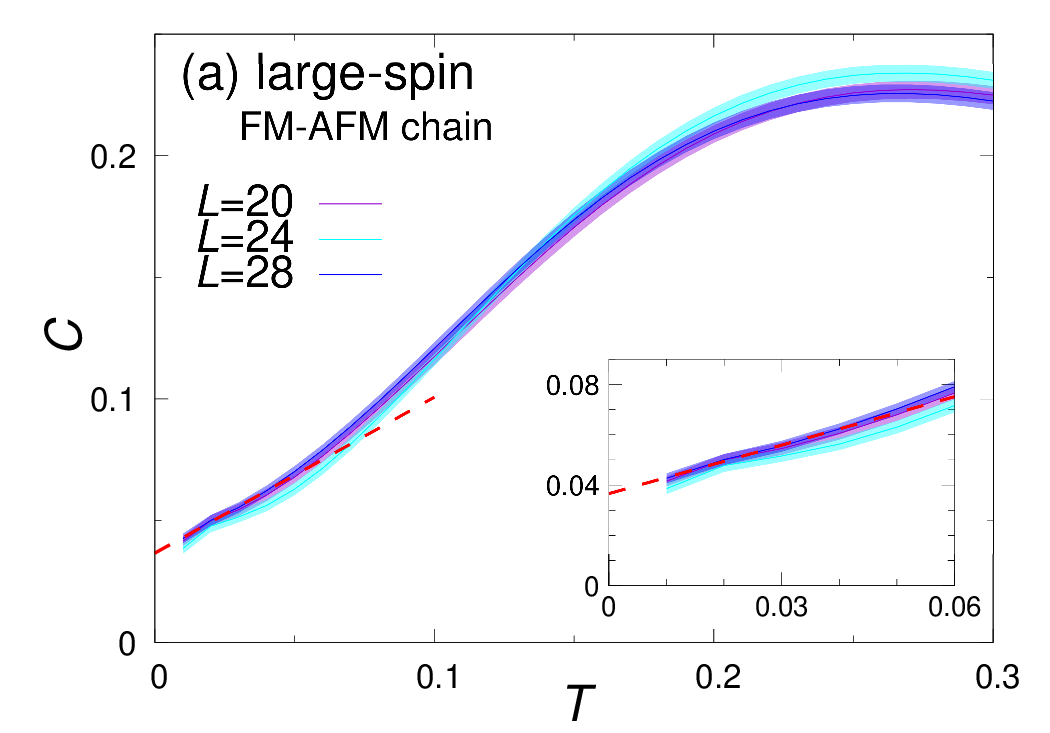}
		\includegraphics[width=8cm]{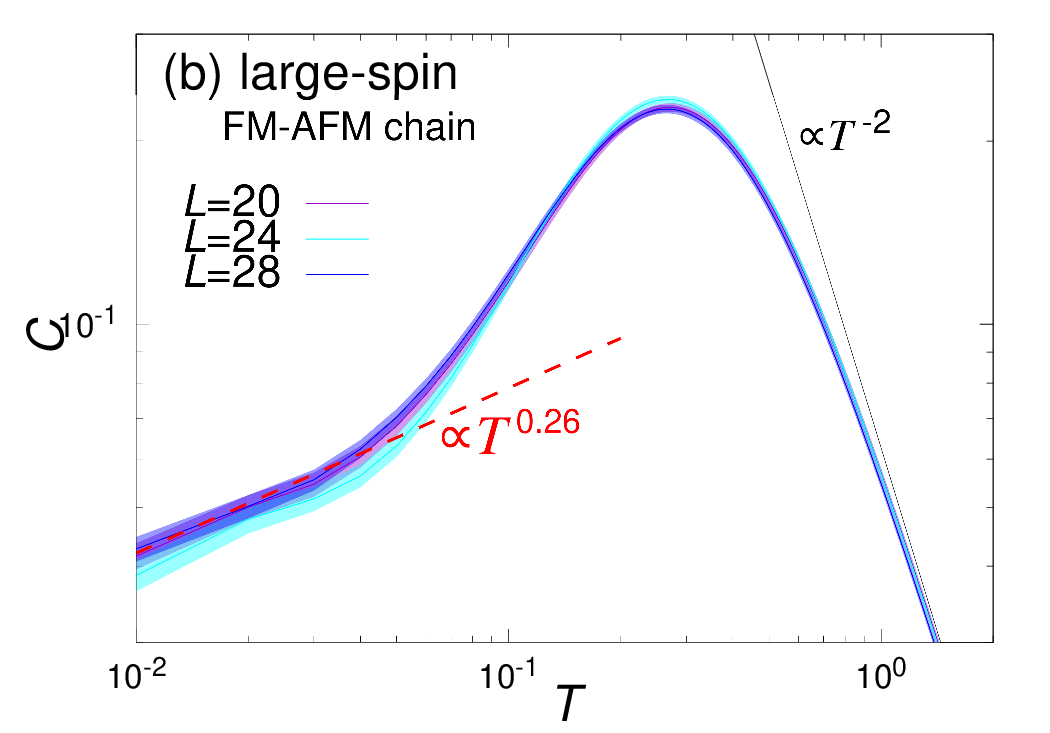}
	\caption{
 (Color online) The temperature and size dependence of the specific heat per spin $C$ of the FM-AFM chain. Figure (a) shows a linear plot, together with a magnified view of the low-$T$ region in the inset, while Fig. (b) shows a log-log plot. The dashed red line in Fig. (a) is a linear fit of the data at $T\leq0.04$, while the dashed red line in Fig. (b) is the power-law fit to the data at $T\leq0.04$. The solid black line in Fig. (b) is the high-$T$ expansion result.
	}
	\label{fig:chainheat2}
\end{figure*}

 For further comparison, we investigate the finite-temperature properties of the random FM-AFM chain. In Fig. \ref{fig:chainheat2}, we show the temperature and size dependence of the specific heat per spin $C$ of the random FM-AFM chain expected to be described by the large-spin FP. As can be seen from the figures, it exhibits the behavior different from the $T$-linear behavior shown in Fig. \ref{fig:chainheat1}. Namely, as shown in the inset of Fig. \ref{fig:chainheat2} (a), the $T$-linear extrapolation yield a spurious $C(T=0)>0$ value, while, as shown in Fig. \ref{fig:chainheat2} (b), the log-log plot yields a slope much smaller than unity, {\it i.e.}, $C\sim T^{0.26}$. The exponent considerably smaller than unity is consistent with the suggestion $C\propto T^{0.44}\log T$ for the large-spin phase \cite{FMSDRG-WesterbergL,FMSDRG-WesterbergB}, though the numerical exponent value is not so close. These results indicate that the observed behavior of the frustrated $J_1$-$J_2$ model shown in Fig. \ref{fig:chainheat1} is significantly different from that of the large-spin phase shown in Fig. \ref{fig:chainheat2}, but quite resembles that of the RS phase in high-$D$ frustrated systems.

 Thus, our conclusion is in contrast to the one suggested from the SDRG studies where properties of both the $J_1$-$J_2$ chain and the FM-AFM chain are characterized by the same large-spin FP \cite{zigzagSDRG-Hoyos,FMSDRG-WesterbergL,FMSDRG-WesterbergB}. Instead, we regard the $R>0$ gapless phase identified in Sect. \ref{sec:GSphase} as the frustrated RS phase, essentially of the same nature as those identified in $d\geq 2$ random frustrated systems in previous works \cite{Watanabe,Kawamura,Shimokawa,Uematsu,Uematsu2,Liu-Sandvik,Wu,Uematsu3,KawamuraUematsu}.

\begin{figure*}
		\raisebox{1cm}{\includegraphics[width=8cm]{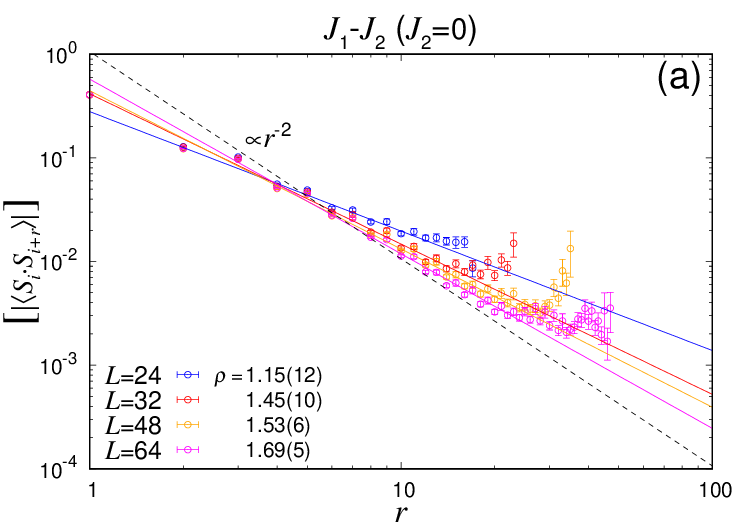}}
		\raisebox{1cm}{\includegraphics[width=8cm]{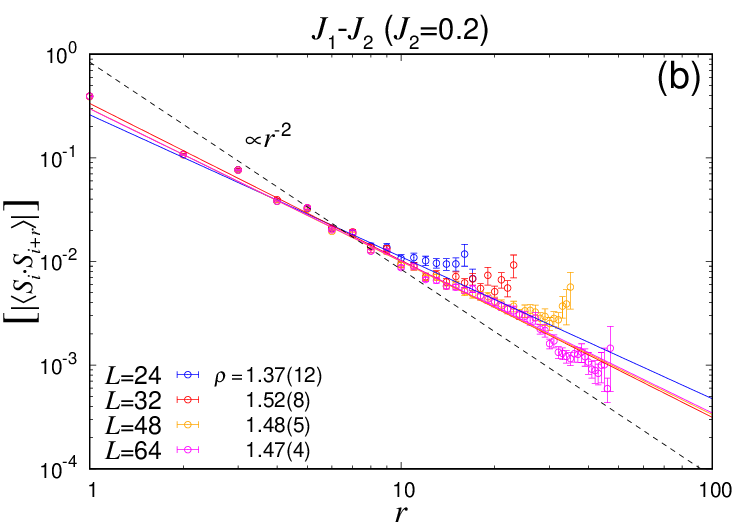}}
		\raisebox{1cm}{\includegraphics[width=8cm]{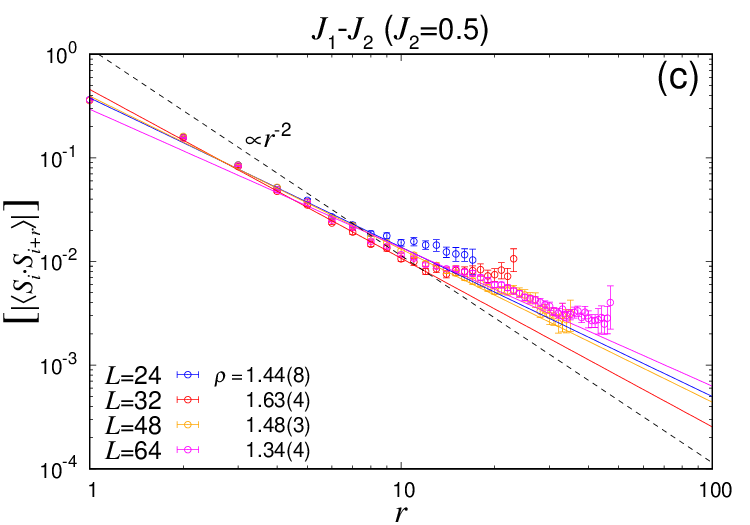}}
		\raisebox{1cm}{\includegraphics[width=8cm]{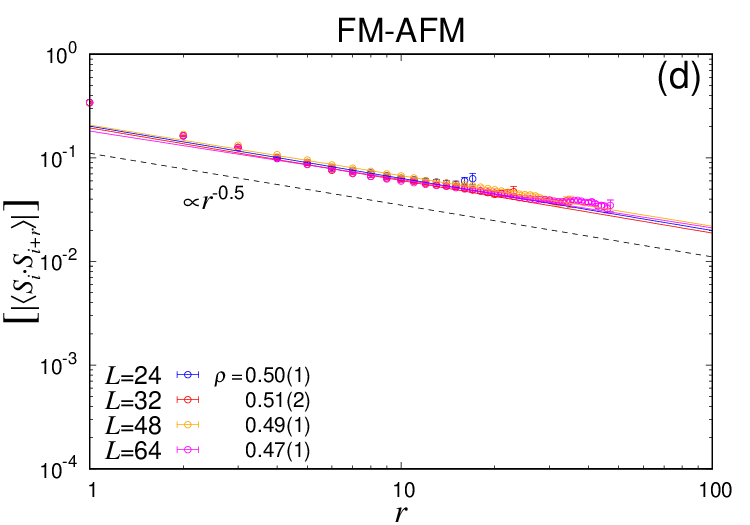}}
	\caption{
(Color online) The log-log plot of the spin-spin correlation function $\left[ |\braket{\bm{S}_{i}\cdot\bm{S}_{i+r}}| \right]$ of (a) the unfrustrated $J_1$-only ($J_2=0$) chain, the frustrated $J_1$-$J_2$ chain of (b) $J_2=0.2$, (c) $J_2=0.5$, and (d) the random FM-AFM chain, for various system sizes as a function of the distance between the spins, $r$. Lines are the power-law fits of the data at $r_{min}\leq r\leq r_{max}$, where $r_{min}=4$ and $r_{max}=\frac{3}{8}L$. The numbers at the legends represent the exponent values $\rho$ describing the power-law decay of the spin-spin correlation functions for each size.
	}
	\label{fig:2PCF}
\end{figure*}

\section{The density-matrix renormalization-group study}
\label{sec:correlation}

 In this section, we present the results of the DMRG study on the ground state of the random $J_1$-$J_2$ chain, including the unfrustrated case of $J_2=0$. We also study the properties of the random FM-AFM chain for comparison. We compute the spin-spin correlation function $\left[ |\braket{\bm{S}_i\cdot\bm{S}_{i+r}}| \right]$ as a function of the distance between spins, $r$.

 Figure \ref{fig:2PCF} exhibits the $r$-dependence of $\left[ |\braket{\bm{S}_i\cdot\bm{S}_{i+r}}| \right]$ on a log-log plot for various system sizes $L\leq64$. (Concerning the convergence problem of the DMRG method, see SIV of supplemental materials \cite{suppl}). In order to get better statistics and lessen the open boundary effects, the $|\braket{\bm{S}_i\cdot\bm{S}_{i+r}}|$ data for a given sample and for a given $r$ are averaged over $\frac{3}{4}L-r$ data points lying in the region $\frac{1}{8}L+1 \leq i < i+r \leq \frac{7}{8}L$. Figure \ref{fig:2PCF}(a) represents the data of the unfrustrated RS state of $J_2=0$, Fig. \ref{fig:2PCF}(b) and (c) the frustrated RS state of $J_2=0.2$ and 0.5, and Fig. \ref{fig:2PCF}(d) the random FM-AFM chain. As can be seen from the figures, when the spin distance $r$ is enough smaller than the system size $L$, say, $r\lesssim \frac{1}{2}L$, $\left[ |\braket{\bm{S}_i\cdot\bm{S}_{i+r}}| \right]$ exhibits a linear dependence on $r$ suggestive of the power-law behavior, while, when $r$ becomes closer to $L$, a significant deviation from the power-law, sometimes even an upward bending, appears for larger $r$ due to the boundary effect associated with the imposed open BC.

We fit the data with the power-law form of $[|\braket{\bm{S}_i\cdot\bm{S}_{i+r}}|]\sim r^{-\rho}$ in the $r$-range of $r_{min}\leq r\leq r_{max}$ to extract the exponent $\rho$. We vary the values of $r_{min}$ and $r_{max}$, to find a rather stable fit arising at around $r_{min}=4$ and $r_{max}=\frac{3}{8}L$. Thus, the fits given in Fig. \ref{fig:2PCF} correspond to $r_{min}=4$ and $r_{max}=\frac{3}{8}L$.

 In the case of the unfrustrated RS state [Fig.\ref{fig:2PCF}(a)], the estimated $\rho$-values monotonically increase as $L$ increases, from $1.15\pm 0.12$ of $L=24$ to $1.69\pm 0.05$ of $L=64$. This observation seems consistent with the relation $\rho=2$ suggested by the SDRG analysis \cite{Fisher,SDRGimprove-Hikihara}. In the case of the frustrated RS state [Figs.\ref{fig:2PCF}(b) and (c)], although the estimated $\rho$-values exhibit a bit more irregular size dependence, most of them distribute around $1.3-1.6$. Hence, while the data are suggestive of the $\rho$-value smaller than two for the frustrated RS state, say, $\rho\sim 1.3-1.6$, we still cannot rule out the possibility of $\rho=2$. We note in passing that the estimated $\rho$-value in the frustrated RS state certainly satisfies the inequality $\rho > \frac{1}{2}$ required to justify the $1/\sqrt{L}$-scaling employed in Sect. III C in the $L\rightarrow \infty$ extrapolation of the spin freezing parameter $\bar q$: Refer to Eq.(1) of supplemental materials SI. \cite{suppl}

 In the large-spin state shown in Fig. \ref{fig:2PCF} (d), the $\rho$-values are even smaller, $\rho \sim 0.5$, and seem to decrease slowly with increasing $L$, down to $\rho=0.47$ for $L$=64. This observation is consistent with the estimate $\rho=0.44$ from the ED data of $\bar{q}$ presented in Fig.5(b). Here, we must mention that we can not exclude the possibility of the decay much slower than the power law, including the logarithmic one \cite{SDRGimprove-Hikihara}. The data for larger $L$ are necessary for determining the asymptotic form of $[|\braket{\bm{S}_i\cdot\bm{S}_{i+r}}|]$ in the large-spin state. Anyway, we may conclude that the decay of spin-spin correlation function in the large-spin state is distinctly slower than that in the frustrated $J_1$-$J_2$ chain of $J_2=0.2$ and 0.5, supporting the view in the preceding section that the ground state stabilized in the frustrating $J_1$-$J_2$ chain with non-negligible $J_2$ is not the large-spin state, but the frustrated RS state.

\section{The nature of the ground state and the low-lying excitation of the frustrated random-singlet state}
\label{sec:excitation}

\begin{figure*}
	\includegraphics[clip,width=8cm]{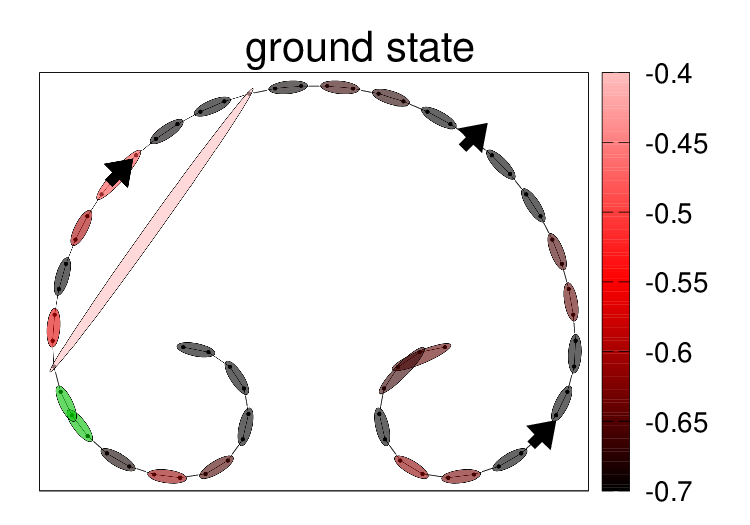}
	\caption{
		(Color online) Typical singlet-dimer configuration of the ground state in a certain sample of the $J_1$-$J_2$ chain of $J_2=0.5$, $\Delta=1$ and $L=64$. Red ellipses represent isolated singlet-dimers, its brightness representing the associated $e_{ij}$-value, {\it i.e.\/}, the $\left <\bm{S}_i\cdot \bm{S}_j\right >$-value given on the bar on the right, while arrows represent orphan spins. Green cluster consisting of more than one singlet-dimers represents the resonating singlet-dimers cluster, a quantum-mechanical superposition of more than one singlet-dimers (and orphan spin) configurations.
	}
	\label{fig:dimer-groundstate}
\end{figure*}
\begin{figure*}
		\includegraphics[clip,width=5cm]{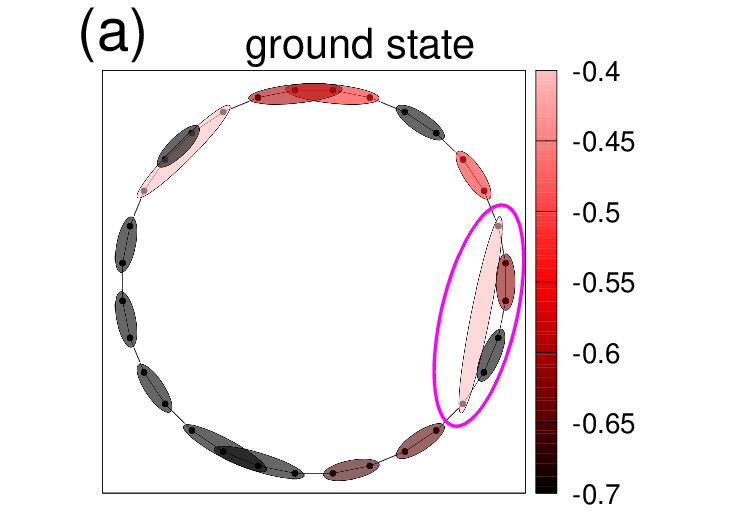}
		\includegraphics[clip,width=5cm]{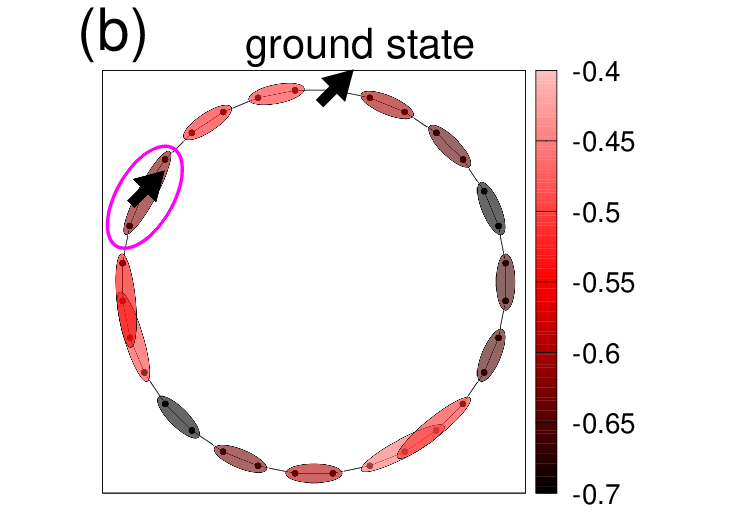}
		\includegraphics[clip,width=5cm]{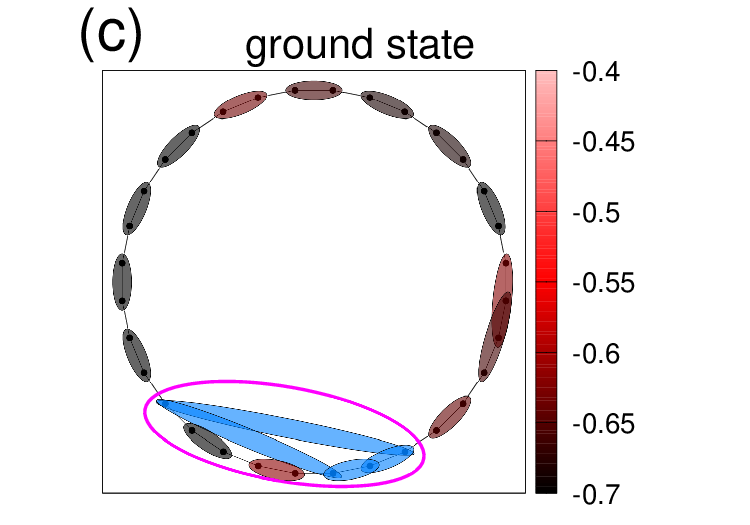}
		\includegraphics[clip,width=5cm]{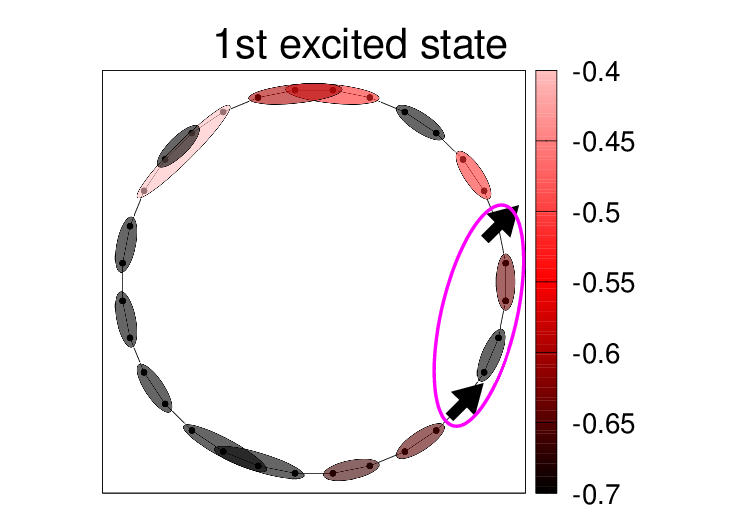}
		\includegraphics[clip,width=5cm]{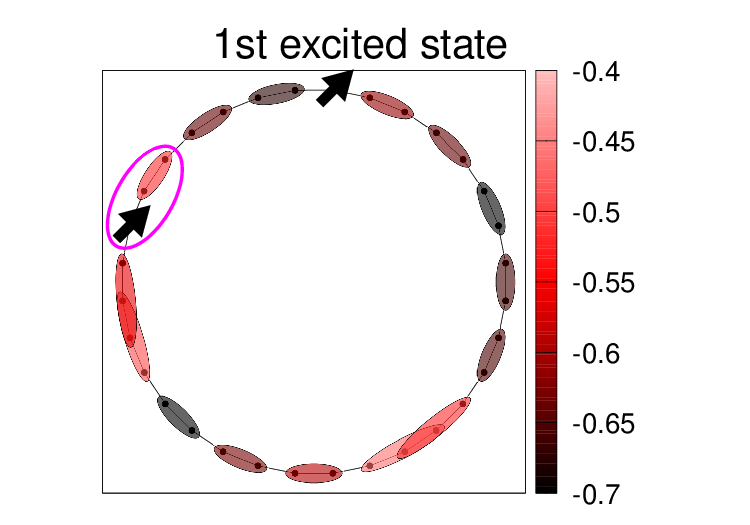}
		\includegraphics[clip,width=5cm]{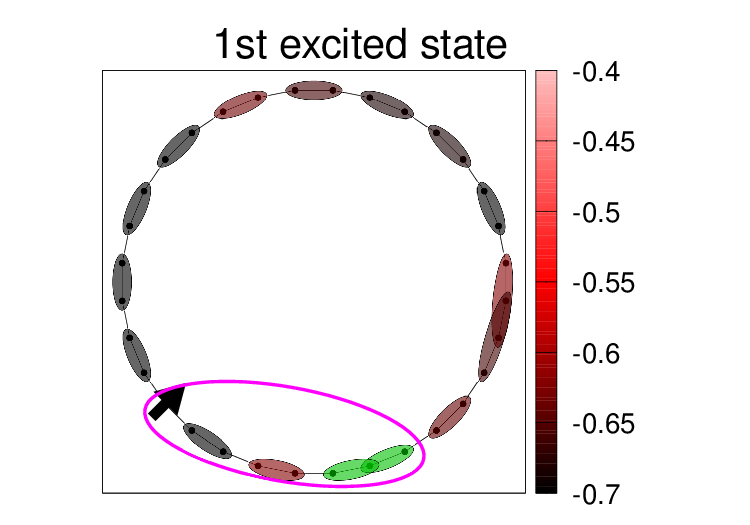}
	\caption{
		(Color online) Typical singlet-dimer configurations of the ground state (upper row) and the first excited state (lower row) in certain samples of the $J_1$-$J_2$ chain with $J_2=0.5$, $\Delta=1$ and $L=32$. Red ellipses represent singlet-dimers with its brightness representing the associated $e_{ij}$-value, while arrows represent orphan spins. Blue (green) clusters consisting of more than one singlet-dimers represent the resonating singlet-dimers cluster. Each low-energy excitation corresponds to (a) the breaking of an isolated singlet-dimer into two orphan spins (singlet-to-triplet excitation), (b) the diffusion of orphan spins accompanied by the recombination of nearby isolated singlet-dimers, and (c) the annihilation of a cluster of resonating singlet-dimers into isolated singlet-dimers and orphan spin.
	}
	\label{fig:configE1excite}
\end{figure*}

In this section, we study the microscopic character of the ground state and the low-energy excitations of the 1D frustrated RS state by investigating the singlet-dimer configurations of the ground state and the first-excited state, following the procedure of Ref.[\citen{KawamuraUematsu}]. In Ref.[\citen{KawamuraUematsu}], the singlet-dimer configurations of the frustrated RS state were studied for several 2D lattices together with those of the 1D unfrustrated RS state in the random $J_1$-only chain. There, it was found that the ground state consisted primarily of the hierarchically-organized singlet-dimers together with the orphan spins and the resonating singlet-dimers clusters, while the ratio of the latter two configurations were suppressed somewhat in the unfrustrated 1D RS state as compared with that of the 2D frustrated RS state.

  In Ref.[\citen{KawamuraUematsu}], the nature of the low-energy excitations were also examined, where the three distinct types of low-energy excitations, labeled as (A), (B), and (C), were identified in the 2D frustrated RS state, {\it i.e.}, (A) the singlet-to-triplet excitation (and its reverse process), (B) the diffusion of orphan spins accompanied by the recombination of nearby singlet-dimers, and (C) the creation (or annihilation) of singlet-dimers clusters. In Ref.[\citen{KawamuraUematsu}], excitations of type (D) were also identified, which are basically variants of type (A). Hence, we regard in the following the type (A) as including the type (D) of Ref.[\citen{KawamuraUematsu}]. Real excitations were generically combinations of these (A)-(C). An interesting observation was that in the 2D frustrated RS state the excitations (B) and (C) dominated, while in the 1D unfrustrated RS state most of the excitations were type (A) with few type (B) or (C). 

 Under such situations, to shed further light on the nature of the possible frustrated RS state in 1D, we examine the singlet-dimer configurations of the ground state and the first excited state of the frustrated $J_1$-$J_2$ chain with $J_2>0$, in comparison with those of the unfrustrated chain with $J_2=0$ and of the frustrated models in 2D \cite{KawamuraUematsu}. 

 In our analysis, all possible bonds (spin pairs) ($ij$) including all distant-neighbor bonds are first ordered according to their two-spin correlation $e_{ij}\equiv \langle {\bm S}_i\cdot {\bm S}_j\rangle$ values in the descending order, {\it i.e.\/}, from smaller ones (negative values with their absolute values large) to larger ones. Then, in the spirit of the SDRG analysis, we draw singlet pairs step by step from the strong bonds to the weaker ones: Namely, the bonds are successively regarded as forming ``singlet-dimers'', under the constraint that a site $i$ which has been involved in already assigned singlet-dimers can no longer be included in a new singlet-dimer. Exception is allowed for the special occasion of the ``local resonance'' where distinct singlet-dimers of ($ij$) and ($ik$) possess nearly degenerate $e_{ij}$ and $e_{ik}$ values (we take the convention of $|e_{ij}-e_{ik}| \leq \frac{1}{32}$), in which case we allow for a ``singlet-dimers cluster'' consisting of more-than-two spins. Such a dimer-formation procedure is repeated until all the remaining spin pairs satisfy the condition $e_{ij}\geq -0.25$ with a vanishing ``entanglement of formation'' (or ``concurrence''), meaning the two spins disentangled. The remaining spins are regarded as ``orphan spins''. For further details of the procedure, refer to Ref.[\citen{KawamuraUematsu}].

In Fig. \ref{fig:dimer-groundstate}, we show singlet-dimer configurations of the typical ground state of the random frustrated $J_1$-$J_2$ chain of $J_2=0.5$ obtained by the DMRG calculation for the $L=64$ chain under open BC. Similarly to those observed in Ref.[\citen{KawamuraUematsu}], the ground state consists of hierarchically organized singlet-dimers, orphan spins, and resonating singlet-dimers clusters. One notable difference from the 2D models is that the singlet-dimers in 1D are not limited to near neighbors but sometimes formed between further neighbors, in contrast to the frustrated RS state in 2D where singlet-dimers are formed primarily between nearest neighbors. The tendency of the singlet-dimers sometimes formed between further neighbors was observed also for the unfrustrated RS state in 1D, meaning such a feature reflects the 1D character \cite{KawamuraUematsu}. 

 In Fig. \ref{fig:configE1excite}, we show singlet-dimer configurations of the typical ground states and the corresponding first excited states of the random frustrated $J_1$-$J_2$ chain of $J_2=0.5$ obtained by the ED calculation for $L=32$ under periodic BC. The comparison of the figures in the upper and lower rows reveals the low-energy excitation of each case.

\begin{figure*}
	\raisebox{1cm}{\includegraphics[width=8cm]{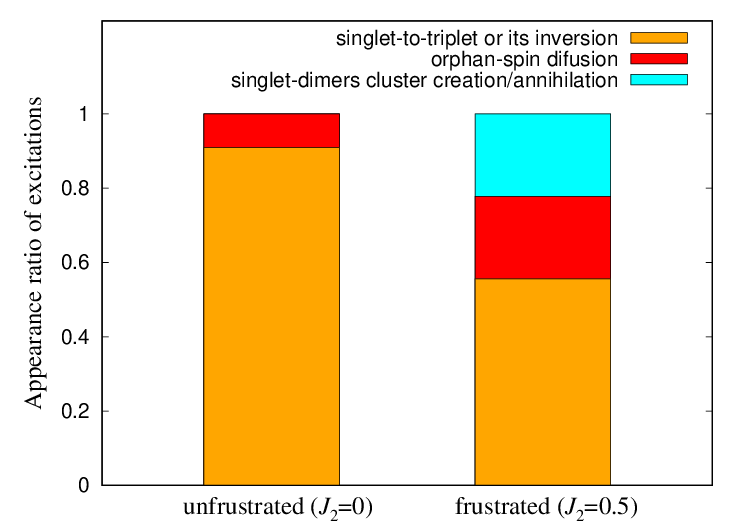}}
	\caption{
		(Color online) The ratio of the appearance probability of different types of excitations in the unfrustrated RS state of $J_2=0$ and the frustrated RS state of $J_2=0.5$, {\it i.e.\/}, `singlet-to-triplet or its inversion' excitation corresponding to type (A) in the text, `orphan-spin diffusion' excitation corresponding to type (B), and `singlet-dimers cluster creation/annihilation' excitation corresponding to type (C).
	}
	\label{fig:ratio-excitations}
\end{figure*}

 As can be seen from these figures, the low-energy excitations consist of the three types (A), (B), and (C) previously identified in  Ref.[\citen{KawamuraUematsu}], each illustrated in Figs. \ref{fig:configE1excite}(a), (b), and (c), respectively. The ratio of the appearance probability of each excitation in the frustrated RS state of $J_2=0.5$ is shown in Fig.\ref{fig:ratio-excitations} in comparison with that in the unfrustrated RS state of $J_2=0$. As can be seen from the figure, in the frustrated RS state, the appearance probability of the types (B) and (C) excitations turns out to be rather high, being of comparable order to that of type (A), in contrast to the unfrustrated RS state of the $J_1$-only chain where the type (A) excitation dominates with rather few (virtually no) type (B) and (C) excitations, demonstrating that the nature of the low-energy excitation of the frustrated RS state of the $J_1$-$J_2$ chain is similar to that of the frustrated RS state in 2D. Such an observation, {\it i.e.\/}, the appearance of considerable amount of type (B) and (C) excitations, which are hardly realized in the unfrustrated RS state but appear in the frustrated RS state in 2D, further strengthens our conclusion that the QSL state in the frustrated random $J_1$-$J_2$ chain is the frustrated RS state, essentially of the same type as stabilized in frustrated 2D and 3D systems.

\section{Summary}
\label{sec:summary}
Both the ground-state and finite-temperature properties of the random-bond $s=1/2$ $J_1$-$J_2$ Heisenberg model on the 1D chain were investigated by means of the ED, DMRG, and Hams--de Raedt methods. The ground-state phase diagram was constructed in the randomness ($\Delta$) versus the frustration ($J_2/J_1$) plane. In the phase diagram, we found two types of randomness-induced states, {\it i.e.}, the unfrustrated RS state and the frustrated RS state. The former is the conventional RS phase discussed in the literature mainly for the 1D unfrustrated random spin chain \cite{DasguptaMa,MaDasguptaHu,Hirsch,Fisher}. The latter, the frustrated RS phase, is essentially equivalent to the one observed in $d\geq 2$-dimensional random frustrated systems, whose existence has not been noticed thus far in 1D systems. It is a phase distinct from the large-spin phase discussed in the SDRG literature \cite{FMSDRG-WesterbergL,FMSDRG-WesterbergB,zigzagSDRG-Hoyos}. Although the reason why the SDRG method yields a qualitatively different answer for the case of the frustrated RS state is not quite clear, we suspect that the important dynamical element of the frustrated RS state, orphan spins which are mobile exhibiting diffusive motion, can hardly be captured within the standard SDRG scheme, and might be the main cause of the failure.

 Our result may suggest that the nature of the frustrated RS state is robust not only to the details of interactions and lattice types but also to the spatial dimensionality including not only 2D and 3D but also 1D, as long as the system possesses a certain amount of frustration and randomness. Together with the results obtained in the previous works \cite{Watanabe,Kawamura,Shimokawa,Uematsu,Uematsu2,Liu-Sandvik,Wu,Uematsu3,KawamuraUematsu}, the frustrated RS state seems to be a highly universal state of quantum magnets, in contrast to the unfrustrated RS state which seems to be rather specific to the 1D unfrustrated system. %and is destabilized by a small amount of frustration or interchain coupling. %\cite{zigzagSDRG-Hoyos}.

\section*{acknowledgements}
This study was supported by JSPS KAKENHI Grants No. 17H06137, 17H02931 and 19K03664. Our code was based on TITPACK Ver.2 coded by H. Nishimori. We are thankful to ISSP, the University of Tokyo, and to YITP, Kyoto University, for providing us with CPU time.
%\end{acknowledgements}

\end{document}